\documentstyle[twocolumn,aps,epsfig] {revtex}
\begin{document}
\bibliographystyle{unsrt}

\title{A Quasi-Classical Model of Intermediate Velocity Particle Production
in Asymmetric Heavy Ion Reactions}

\author{A. Chernomoretz$^1$,
L. Gingras$^2$,
Y. Larochelle$^2$\thanks{Present address: Joint Institute for Heavy Ion Research,
Holifield Radioactive Ion Beam Facility, Oak Ridge, Tennessee 37831-6374.},
L. Beaulieu$^2$,
R. Roy$^2$, C. St-Pierre$^2$, C.O.Dorso $^1$} 
\address{$^1$ Departamento de Fisica, Universidad de Buenos Aires,
Buenos Aires, Argentina.}
\address{$^2$ Laboratoire de Physique Nucl\'eaire, D\'epartement de
Physique, Universit\'e Laval, Qu\'ebec, QC, Canada G1K 7P4.}

\date{\today}
\maketitle
\begin{abstract}
The particle emission at intermediate velocities in mass asymmetric 
reactions is studied within the framework of classical molecular
dynamics. Two reactions in the Fermi energy domain were modeled 
, $^{58}$Ni+C and $^{58}$Ni+Au at 34.5 MeV/nucleon.
The availability of microscopic correlations at all times allowed a
detailed
study of the fragment formation process. Special attention was paid to
 the physical origin of fragments and emission timescales, which allowed us to 
disentangle the different processes involved in the mid-rapidity particle
production. Consequently, a clear distinction between a prompt pre-equilibrium
emission and a delayed aligned asymmetric breakup of the heavier partner
of the reaction was achieved.
\end{abstract}
PACS number(s): 25.70 -z, 25.70.Mn, 25.70.Pq, 02.70.Ns 

\section{Introduction}

Heavy ion reactions in the Fermi energy domain (20 - 100 MeV/nucleon) has
been an intense field of research during the last decade.
In particular, a lot of attention was paid to the strong emission
of intermediate mass fragments (IMF) observed in the
mid-rapidity range.
\cite{mont94,leco95,toke95,demp96,laro97b,luka97,lefo00,plag00,poggi01}.
Lying in rapidity space between the contributions
of the projectile-like and target-like fragments,
this complex emission pattern can not be explained as a result
of a simple statistical emission of the two excited participant nuclei. 
This fact suggests that it should be considered as a non-trivial
phenomenon in which the interplay of the two-body interactions and
the nuclear mean field plays a crucial role. 

There is not yet a uniform agreement about the possible mechanisms
responsible for this IMF overproduction at mid-rapidity.
Recently, a statistical approach to the problem was developed
\cite{stat}, and the claim that the inclusion of Coulombian interactions
in the statistical picture could explain this behavior was made. On the contrary,
some authors adopted a dynamical description to analyze the same process. 
For example, a Landau-Vlasov semiclassical transport model was used to analyze 
the presence \cite{luka97} and the temporal behavior \cite{eudes00} of
the particle emission process that populates the mid-rapidity range. 
Neck instabilities, dynamical fluctuations, and particle production at
mid-velocities were also studied using the quantum molecular dynamics (QMD)
model in semi-peripheral \cite{colonna95}, and central 
heavy ion collisions \cite{nebauer99}.
In addition, in a recent contribution, \cite{pian01}, two mechanisms of IMF
production were suggested (neck, and fission-like surface emission), and
two different models were introduced in order to describe them.

The aim of this paper is to present a simple model that
provides a description of the IMF production process at
mid-velocities. In particular, we focus our attention on two mass asymmetric
reactions of relatively small systems, 
$^{58}$Ni+C and $^{58}$Ni+Au at 34.5 MeV/nucleon. The motivation to
study these systems is two-fold. On one hand, it allows us to resolve the
controversy about fast overlap emission processes and slow surface deformation
breakup.  On the other, our time-based analysis supports new experimental
observations of alignment and proximity effects in neck breakup of mass
asymmetric collision partners \cite{ging01}.

The reactions are studied within a quasi-classical approach using molecular
dynamics techniques.
Out of various methods used to study heavy ion reactions, the implemented MD model
can describe changes of phase, hydrodynamic flow, and non-equilibrium
dynamics without adjustable parameters. All order nucleon-nucleon
correlations to form fragments and neck-like structures are
intrinsically incorporated in the dynamics. Of course, the price one has to pay
using such a simplified  quasi-classical description is that 
quantitative agreement with experiments can not be taken for granted. Nevertheless, 
despite its simplicity, we found it a great tool to probe new physical effects and
to propose a meaningful scenario for different origins of MR particles
(see \cite{ging01}).

Several kinematical features of the experimental data are well reproduced. 
In addition, the availability of correlations of
all orders at all times allowed us to study in detail the
time evolution of the fragment formation
process of IMF's and to identify different mechanisms of mid-rapidity particle production
within a unified description. In this way, three time-scales
could be differentiated: a violent pre-equilibrium stage, a delayed aligne
emission stage, and a ``statistical" evaporation stage.

In section II, we briefly describe the experimental setup. The description of the
nuclear interactions model and the used fragment recognition algorithm are given
in sections III and IV respectively. The obtained results and their
posterior analysis are included in sections V and VI. Finally, conclusions are drawn 
in section VII.

\section{Experimental Data}
The two mass asymmetric reactions studied in this contribution 
were performed at the coupled Tandem and Super-Conducting
Cyclotron accelerators of AECL at Chalk River.
A beam of $^{58}$Ni accelerated at 34.5 MeV/nucleon bombarded
alternatively a
2.4 mg/cm$^{2}$ carbon target and a 2.7 mg/cm$^{2}$ gold target.
Charged particles issued from these reactions were detected in the
CRL-Laval 4$\pi$ array
constituted by 144 detectors set in ten rings concentric to the beam
axis and covering polar angles between 3.3$^o$ and 140$^o$.
See \cite{ging98a} and references therein for more information on
detectors and energy calibration.
In the present work, we will consider events selected in the off-line
analysis by a total detected charge of at least 24 and 34 (fully
detected) units for the
 $^{58}$Ni+Au, and $^{58}$Ni+C systems respectively.

\section{The Model}\label{model}
In order to get information on the time formation and the kinematical 
characteristics of the intermediate
velocity material produced in these mass asymmetric heavy ion reactions, we
relied on molecular dynamics simulations.
The model used in this work was originally introduced by Lenk {\it et al.} 
to test the accuracy of the Vlasov-Nordheim 
approximation~\cite{pandha}. It is a classical molecular dynamics 
(CMD) model with a spherical two-body interaction potential in which 
Coulombian interactions are taken into account, while the nuclear flavor is
provided by the following interaction terms:
\begin{eqnarray}
V_{np}&=&V_r [{e^{-\mu_r r} \over r} - {e^{-\mu_r r_c} \over r_c}] - V_a
[{e^{-\mu_a r} \over r} - {e^{-\mu_a r_c} \over r_c}] \\
V_{nn}&=&V_0 [{e^{-\mu_0 r} \over r} - {e^{-\mu_0 r_c} \over r_c}]
\end{eqnarray}
 $V_{np}$ and $V_{nn}$ stand for neutron-proton and
identical nucleons interactions respectively, and $r_c=5.4$ fm is a cutoff 
radius. We used a set of parameters (M-set in Ref.~\cite{pandha}), that
gives an equation of state with a compressibility around $250 MeV$,
an equilibrium density of $\rho_0=0.16 fm^{-3}$, and $E(\rho_0)= -16 MeVA$.
Due to the particular choice of parameters, this model provides a realistic
nucleon-nucleon cross section and it has been used to reproduce reasonably
well several features of experimental data on heavy ion collisions (see ~\cite{pandha,cmd1,cmd1b,cmd2} and references therein)

The nuclei used in our computational experiments have been
built with the right number of protons and neutrons while the corresponding 
ground-states were obtained by molecular dynamic techniques, i.e. by a cooling
procedure that starts with a confined and rather excited nucleus, and ends 
with a self-contained state at reasonable binding energies. 
Before each collision a random relative projectile-target orientation was
chosen, and afterwards the projectile's center of mass velocity 
was boosted to the desired energy.
The set of equation of motion is integrated up to a final time of
$t=2000$ fm/c with a standard velocity-Verlet algorithm~\cite{verlet},
taking $t_{int}=0.02$ fm/c as the integration time
step, achieving an energy conservation of $0.01\%$. Under this scheme, we
have analyzed $8400$ collisional events for the $^{58}$Ni+$^{12}$C at $34.5$ MeV/nucleon reaction, for a wide range of impact parameter values 
($b \le 4.5$ fm), and $3400$ events for
the $^{58}$Ni+$^{197}$Au at $34.5$ MeV/nucleon case ($5 $fm$ \le b \le8 $fm).

\section{Fragment Recognition}\label{fragment}
In order to explore the nucleon-nucleon correlations that give rise to the 
fragment formation process, we have adopted a definition of {\em cluster} 
compatible with the clusterization algorithm known in the literature
as the {\it minimum spanning tree in energy space} (MSTE)~\cite{mste}. 
Under the MSTE scheme a given set of particles
$i, j,..., k$ belongs to the same cluster $C_a$ if:

\begin{equation}\forall \, i \, \in \, C_a \:,\: \exists \, j
\, \in \, C_a  \, / \,  e_{ij} \leq 0
\end{equation}
where $e_{ij} = V(r_{ij}) + ({\bf p}_i - {\bf p}_j)^2 / 4 \mu$,
and $\mu$ is the reduced mass of the pair $\{i,j\}$. 
This clusterization method  searches for correlated structures
in {\bf q} space (first term in the definition of $e_{ij}$),
with the additional possibility of avoiding 
certain particle incorporation to a given cluster regarding at the relative 
momenta of particle pairs (second term of $e_{ij}$). 
In this way the conformation of the MSTE partitions reflects 
 certain degree of correlation in {\bf q}-{\bf p} space.
Due to this feature, the MSTE becomes very useful in the
recognition process of promptly emitted particles during the  pre-equilibrium 
stage of the reaction.

The availability of correlations of all-orders at all times allows
us to gain a lot of insight about the presence and origin of mid-velocity
sources. For instance, right after the most violent stage of the collision,
 an MSTE cluster recognition 
step enables the tagging of particles as belonging to the quasi-projectile 
(QP), the quasi-target (QT), or as free particles (FP) not 
{\bf q}-{\bf p} correlated neither to the QT nor the QP at
{\it tagging time} ($t_{tag}$). 

\begin{figure}
\centerline{\epsfig{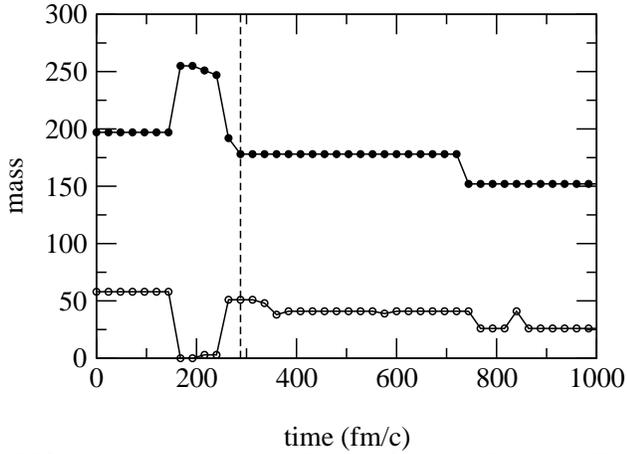}}
\caption{Mass temporal evolution of the two biggest MSTE clusters 
for a typical $^{58}Ni+^{197}Au$ reaction with an impact parameter value of
$b=6 fm$. The dashed line signals the estimated $t_{tag}$. \label{fig_ttag} }
\end{figure}

The value of $t_{tag}$ is determined on an event by event basis.
Analyzing the temporal behavior of the size of the two biggest clusters,  
$t_{tag}$ is associated with the time at which the mass number of these
fragments attains stability for the first time after the
fragmentation process has started ~\cite{aclara}. 
Figure ~\ref{fig_ttag} illustrates this last point.
It shows a typical $t_{tag}$ determination for a
single $^{58}Ni+^{197}Au$ reaction with an impact parameter value of
$b=6 fm$.
 
It is then clear that $t_{tag}$ signals the entrance into a new and less violent
deexcitation regime
with particles QP(QT)-tagged being well correlated 
in configuration space right after the highly collisional stage. 

In addition, the microscopic description allows to easily calculate the time 
 ($t_e$) at which every fragment has actually been
emitted by simply tracing back the dynamical evolution of the reaction. Since each
 asymptotic detected cluster is emitted at different stages of the reaction 
 (a cluster is considered as a whole when most of its nucleon
 constituents get well correlated velocities and positions,
 differentiating themselves from other fragments)
it is possible to extract useful information
 correlating this {\it emission time} characterization with other
kinematical variables.

\section{Results}\label{results}
\subsection{Reliability of the simulations}\label{simulation}
In order to compare the numerical simulations with the experimental data 
the following prescription was adopted to estimate an observable
($b_{exp}$) related to the reaction impact parameter 
(see~\cite{ging01},\cite{ging98b}):
\begin{equation}
b_{exp}=r_0 (A_P^{1/3} + A_T^{1/3}) \frac{\Pi_{||}^{cm}}{P_P^{cm}}, 
\end{equation} 
 where $\Pi_{||}^{cm}$ is the total parallel momentum of all charged
particles in the forward velocity hemisphere of the center of mass (CM) reference
 frame, $A_P$, $A_T$ and $P_P^{cm}$ are respectively the projectile
mass number, target mass number and projectile CM momentum. We took 
$r_0 = 1.2 fm$.

\begin{figure}
\centerline{\epsfig{figure=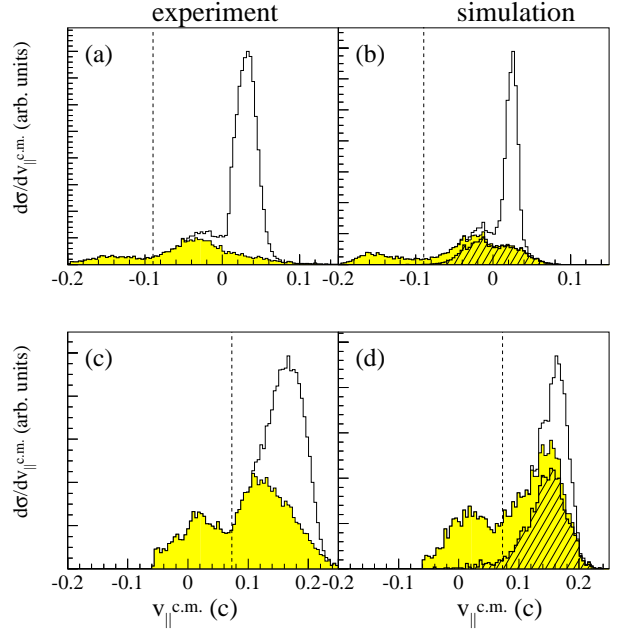,height=9cm}}
\caption{ Normalized parallel velocity distribution, calculated in the
center of mass system for $z \ge 3$ particles, corresponding to mid-peripheral and
peripheral $^{58}Ni+C$ and $^{58}Ni+Au$ reactions are shown in upper and lower
panels respectively. Vertical dashed lines were drawn at nucleon-nucleon 
velocity value. Figures (a) and (c) show experimental results, whereas (b) and (d) show simulated ones. In each panel the IMF contribution is shadowed, whereas QP tagged IMF's are stripped. See text for details.
 \label{fig_compara} }
\end{figure} 

In panels (a) and (b) of figure~\ref{fig_compara} we show the normalized 
parallel velocity distribution ($v_{||}^{cm}$), in the CM reference frame, of 
the $z \ge 3$ collision charged products for $^{58}$Ni+$^{12}$C peripheral and
mid-peripheral ($b_{exp} \ge 2.5$ fm) reactions. Experimental data is
shown in the first panel and simulation results in the second one.
The IMF contribution($3 \le z \le 7$) was shadowed and quasi projectile
assigned IMF were stripped . 
(The 4$\pi$ detector coverage and the complete detected charge condition 
allow us to establish, in this case, a meaningful comparison between the
experimental data and unfiltered simulated reactions ).

It can be seen that general features are well reproduced by the
simulation. Aside from PLF and TLF velocity signatures, an intermediate
bump peaked in between $v_{nn}$ and PLF velocity can be recognized for
both, experimental and simulated data. 
Nevertheless, it seems that the CMD model overestimates slightly the
amount of IMF's coming from the QP in the higher velocity region of
$v_{||}^{cm} \sim 0.03 c$.

The corresponding normalized parallel velocity distributions 
 for $^{58}Ni+Au$ mid-peripheral and peripheral reactions ($6$ fm $< b_{exp}$),
are shown in panels (c) and (d) of figure \ref{fig_compara}. In this case, simulated
data were filtered using a software implementation of the detector.
Despite the completely different entrance channel of this reaction, and the
different effect played by the detection thresholds of the apparatus, similar
conclusions can be drawn for this reaction: general features of the
distribution are well reproduced by the model, but an overproduction of
IMF coming from the QP (centered at $v_{||}^{cm} \sim 0.15 c$) can also
be recognized in panel (d).

We believe that this 
happens because our simplified model fails to accurately reproduce secondary deexcitation
processes at a quantitative level. In particular, the relative amplitude of LCP's and
IMF's production in the QP exit channel is not well reproduced
because the quasi-classical approach can not adequately handle the relevant degrees
of freedom involved in the slow nuclear deexcitation processes. 

Another point that is worth noting is that, as the difference between 
the mid-rapidity bump (on the QP side) and $v_{nn}$ velocities seems
to be similar for both reactions (panels (a), and (c)),
one can conclude that the deexcitation process of the QP should
be the same.  Within our model, we can not rule out the possibility that in both reactions the
QP proceed thru similar IMF production mechanisms. In particular, it seems
possible that the process of delayed asymmetric breakup (see Sec.\ref{time}) is there
in both cases. However, along the present contribution, plausibility arguments will be given
supporting the idea that this happens with different probabilities in each case.

From the preceeding analysis, it can be concluded that, 
even if a quantitative description
of the process seems to be questionable, the use of our CMD model
can throw some light upon the problem of mid rapidity production. As will be shown, despite 
its simplicity, the model is flexible enough to allow the presence of
qualitatively different mechanisms of particle production that can be
integrated in a unified description of the process. This can be used to
settle a meaningful scenario within which experimental data can be
studied.

\subsection{CMD Description}
\subsubsection{Source-origin analysis}
The microscopic description
of the reaction allows us to associate two parameters to each 
asymptotic charged particle: its emission time ($t_e$), and its early 
physical origin, i.e if the particle belonged to the
quasi-projectile(QP),  quasi-target (QT) or was a prompt free particle (FP)
at $t_{tag}$. 

\begin{figure}
\centerline{ \epsfig{figure=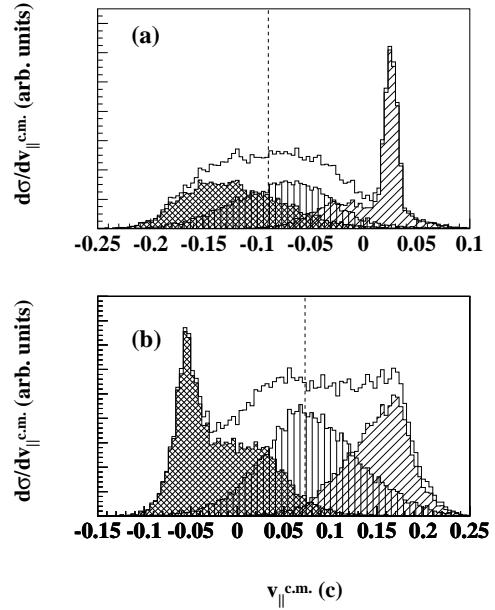,height=9cm}}
\caption{Normalized parallel velocity distribution, in the system center of mass,
for every emitted charged particle in mid-peripheral $^{58}$Ni+$^{12}$C (a), and
 $^{58}$Ni+$^{197}$Au (b) simulated reactions.
The dashed vertical line corresponds to the nucleon-nucleon velocity value.
The contribution of QP, FP, and QT-tagged particles are plotted 
striped, shaded and grided respectively.\label{fig_simnicniau} }
\end{figure}

In figure~\ref{fig_simnicniau} we show the asymptotic distribution of 
$v_{||}^{cm}$ calculated for every charged particle, for the two analyzed reactions.

We considered, as in figure~\ref{fig_compara},
mid-peripheral and peripheral events. 
In the figures, the contribution of particles that had
been labelled as QT, FP, and QP at $t_{tag}$, are distinctively shadowed.

Figure~\ref{fig_simnicniau}~(a) corresponds to the
$^{58}$Ni+$^{12}$C reaction ($2.5$fm$\le b_{exp} < 6.0$fm). 
It can be seen that the particles tagged as FP were mainly emitted slightly above $v_{nn}$.
Comparing with figure~\ref{fig_compara}~(b), it can be noticed
that this emission is almost entirely composed by LCP's, dynamically
expelled in the early stage of the reaction. 
The wide velocity distribution of the QT-tagged particles 
reflects the fact that the $^{12}$C target is easily broken in the collision
producing mainly $z\le 3$ as byproducts. 
Finally it can also be seen, for the QP-tagged particles, an important
contribution of intermediate mass fragments and heavier particles, aside from 
the presence of LCP's (see figure~\ref{fig_compara}~(b)).
In this case, as was remarked earlier, not only a maximum centered at the
projectile-like fragment (PLF) velocity can be seen, but a second peak can also
be recognized towards the mid-velocity regions. 

Looking at figure~\ref{fig_simnicniau}~(b), a similar analysis can be
performed for $^{58}$Ni+$^{197}$Au collisions ($9.0$ fm$\le b_{exp} < 12.0$ fm).
In this case the FP 
parallel velocity distribution is centered at $v_{nn}$.  Again this kind of emission
is mainly composed of promptly emitted LCP's.
The QP emission achieves its maximum at the PLF
velocity, 
and presents a slight asymmetry towards the mid-velocity region. Indeed, this
last feature is much more noticeable on the side of the heavier partner of the
reaction, where an extra contribution towards the mid-rapidity, aside of
the parallel velocity distribution centered around the target-like fragment (TLF)
velocity, is easily recognizable.

\subsubsection{IMF's emission-time analysis}\label{time}

As we mentioned at the end of section~\ref{fragment}, in our CMD simulations, the
emission time ($t_e$) of asymptotic clusters can be easily estimated.
At this point, it is important to remember that the tagged-origin 
classification reflects spatial and velocity correlations right after the compound nucleus starts to fragment,
while $t_e$ signals the time when a given cluster attains mass stability.

\begin{figure}
\centerline{\epsfig{figure=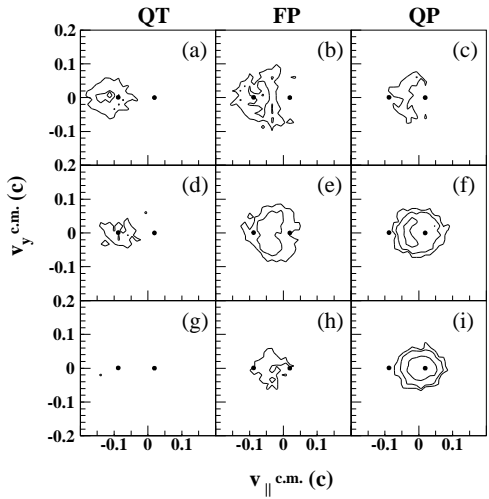,height=9cm}}
\caption{ Two-dimensional velocity distribution,  for mid-peripheral
$^{58}$Ni+$^{12}$C simulated reactions. A cut perpendicular to the
reaction plane is shown, i.e. ($v_y$,$v_z$). 
The first, second, and third
columns include data for QT, FP, and QP-tagged IMF's respectively.
The first, second and third figure rows show
the contribution of particles emitted at $t_e <150$ fm/c, $150$ fm/c$\le t_e < 600$ fm/c, 
and $t_e \ge 600$ fm/c after the beginning of the collision.\label{fig_nicteyz}}
\end{figure}

In figure~\ref{fig_nicteyz} we used such characterization in order to plot
2-D velocity distributions, in the CM system, of IMF particles ($3\le Z \le 7$) for  
$^{58}$Ni+$^{12}$C mid-peripheral reactions ($2$fm $\le b_{exp} \le 4$fm). 
The figures show the velocity distribution in the perpendicular plane of
the collision ($v_y-v_z$ plane), and a logarithmic scale was used for the
contour levels.  
The contribution of QT, FP, and QP-tagged particles are shown in the first,
second, and third column respectively, while particles emitted at times 
$t_e < 150$ fm/c, $150$ fm/c$ \le t_e < 600$ fm/c, and
$t_e \ge 600$ fm/c after the collision are shown in the first, second and
third row respectively.  
In the figures, we included, as reference, the mean velocities
(taken over all charged particles) of QP-tagged and QT-tagged particles (solid circles).

It can be seen that the QT-tagged IMF particles are emitted very early in the
evolution (Figure~\ref{fig_nicteyz}~(a)).
 They correspond mainly to small IMF's that come from the $^{12}$C breakup 
and, eventually, some dragged $^{58}$Ni nucleons.
On the other hand the FP-tagged particles (middle panels of figure
\ref{fig_nicteyz}), that were already differentiated at $t_e$ 
from the quasi-projectile and the quasi-target, seem to be emitted
in a {\it conical} pattern for $t_e < 150$ fm/c reflecting the violence of the collision
and the extremely asymmetric entrance channel.
 A second contribution of this kind of particles occurs a little
bit later, $150$ fm/c$ \le t_e < 600$ fm/c, when excited promptly ejected
FP-clusters emit a second wave of FP-fragments that 
populate the intermediate velocity region, see Figure~\ref{fig_nicteyz}~(e).
(It is important to keep in mind, though, that the main contribution of
FP-tagged particles are not IMF's but LCP's)
Finally, the QP-tagged particles, shown in the third column, exhibit a very
interesting behavior. The maximum IMF emission of this kind of particles occurs
at intermediate times, $150$ fm/c$ \le t_e < 600$ fm/c, after the collision
(Figure~\ref{fig_nicteyz}~(f)).
 A Coulomb hole in the emission pattern, that reflects the interaction between the 
QP-residue and the emitted IMF clusters, can easily be recognized. This
kind of pattern explains the second maximum observed in the parallel velocity
distribution shown in Figure~\ref{fig_simnicniau}(a), and can be considered as
the signature of a dynamically induced {\it delayed QP breakup}.
Note also that after this kind of delayed emission, a subsequent
isotropical pattern can be observed for longer times (Figure~\ref{fig_nicteyz}~(i)).

\begin{figure}
\centerline{\epsfig{figure=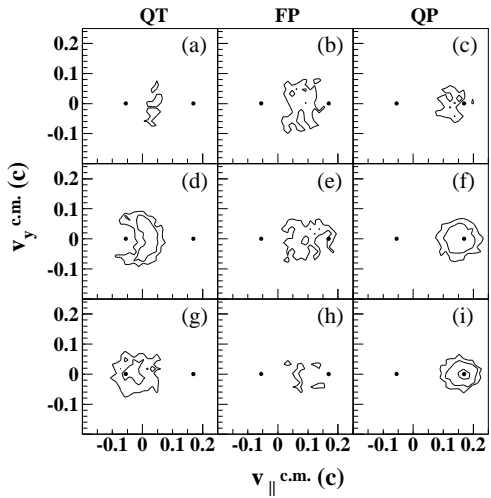,height=9cm}}
\caption{Same as figure 4, but calculated for $^{58}$Ni+$^{197}$Au simulated reactions.\label{fig_niauteyz} }
\end{figure} 

The same analysis can be applied to the $^{58}$Ni$+^{197}$Au reaction (see 
Figure~\ref{fig_niauteyz}), where the same kind of behavior can be observed,
for very short and very long emission times (first and third row
respectively).
Nevertheless, the emission pattern displayed by the system at intermediate times,
$150$ fm/c$ \le t_e < 600$ fm/c, after the collision presents a special interest.
One can observe that in this case, a {\it delayed} IMF emission pattern is
recognizable mainly on the QT side of the reaction, while the QP related IMF
emission is nearly isotropic at this times, in contrast to figure
\ref{fig_nicteyz}(f).

\begin{figure}
\centerline{\epsfig{figure=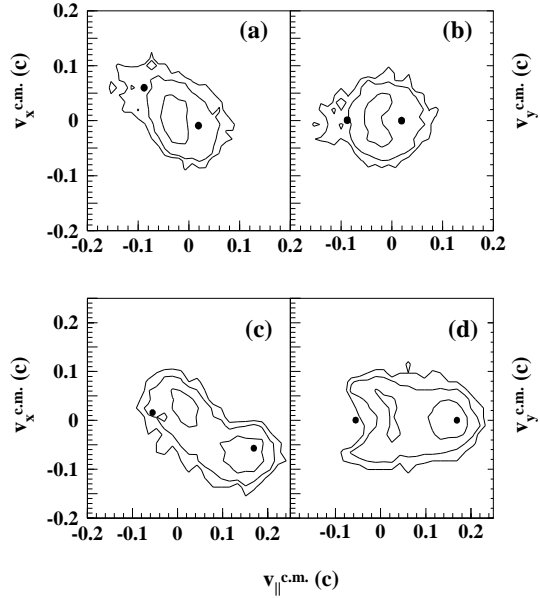,height=9cm}}
\caption{Two-dimensional velocity distributions for IMF particles emitted
between $t_e=150$fm/c and $t_e=600$fm/c. Frames (a) and (b) correspond
to mid-peripheral $^{58}$Ni+$^{12}$C simulated reactions, while frames
(c) and (d) to $^{58}$Ni+$^{197}$Au ones. The left column shows
 the reaction velocity plane, ($v_x,v_z$), while the right 
column shows a perpendicular cut ($v_y,v_z$).\label{fig_xzyz} }
\end{figure}

We complete the characterization of the {\it delayed} IMF emission observed in
both reactions with figure~\ref{fig_xzyz}. 
This figure shows, for $150$ fm/c$ \le t_e < 600$ fm/c, the two-dimensional velocity
distributions in the reaction plane (first column), and in the perpendicular plane
(second column). The first and second rows of figures correspond to
$^{58}$Ni$+^{12}$C and $^{58}$Ni$+^{197}$Au respectively. For the sake of clarity, 
we did not make any distinction between QP, QT or FP-tagged contributions in this case. 
It can be seen that the observed delayed IMF emission occurs mainly along the
QP-QT direction in both reactions. A slight deviation of this alignment is
observed due to the angular momentum gained by the emission sources after the
collision.

\subsubsection{Role of Coulomb Interactions \label{sec_coulomb}}

In this section we will analyze the specific role played
by the Coulombian interactions in the detected delayed IMF
emission. To that end we study the $^{58}$Ni + Au reaction, comparing
the behavior of the system when Coulomb interactions are not considered.

\begin{figure}
\centerline{\epsfig{figure=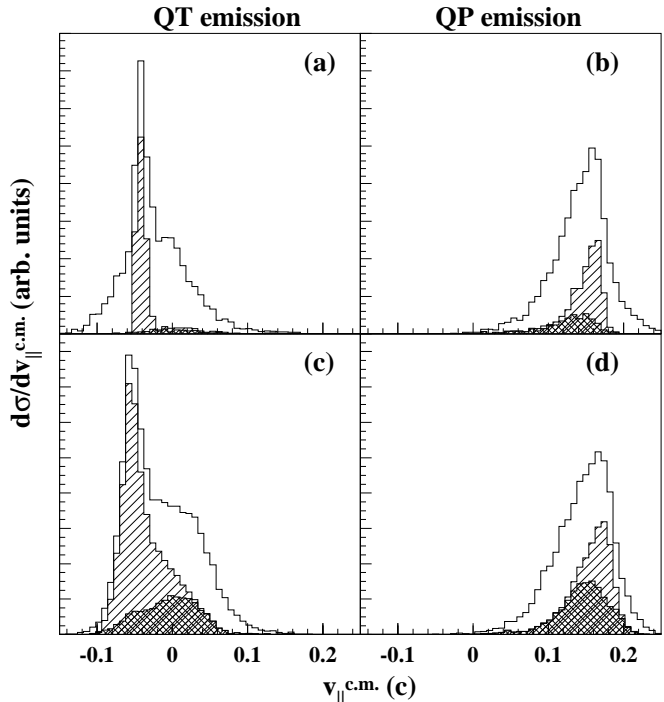,height=10cm}}
\caption{Parallel velocity distribution of QT and QP emitted particles
(panels (a),(c), and (b),(d)) in the simulated $^{58}$Ni+Au reaction.
$z\ge3$ particle contribution is stripped, whereas IMF contribution is
cross-hatched. The upper panels show simulations without Colombian
interaction. The lower ones include it.
\label{fig_coulomb} }
\end{figure}

In figure \ref{fig_coulomb} parallel velocity distributions are shown
when Coulomb is neglected (panels (a) and (b)) and when it is considered
in the simulated evolutions
(panels (c) and (d)). QT emitted particles and QP emitted ones are
displayed on the first and second columns respectively. The contribution
of $z\ge3$ particles is stripped, whereas IMF contribution is
cross-hatched.
 
It can be seen from the figure that the QT emission pattern is
qualitatively affected by the presence of Coulombian interactions.
In panel (a) the QT IMF production is negligeable. The few IMF
present in that picture are barely attached to the quasi target
, being realeased within a very short time after the projectile pass
through. Without Coulomb instabilities, the dynamically induced deformations
are mostly reabsorbed by the TLF. 

This is not the case when electrostatic interactions
are considered (panel (c)). Moreover, the overall QT emission pattern in panel (a)
is rather isotropic and have lost, up to a certain degree, the memory of the entrance channel.
On the contrary, the QT emission pattern when Coulombian interactions
are included in the Hamiltonian (panel (c)), is severely affected by the 
collisional dynamics, giving place to what we called {\em IMF aligned delayed} emission.

The QP emission pattern, on the other hand, shows the same qualitative
behavior with or without Coulomb. That would imply that electrostatic
interactions are not the key ingredient in the QP deexcitation
process.

\section{Discussion}
All of these findings show the existence of two different scenarios 
for the origin of intermediate velocity particles.
On one hand, there is an important prompt emission of LCP particles,
at the highly collisional stage of the reaction. They come mainly from the overlap zone,
as a consequence of nucleon-nucleon collisions that promptly eject them from the
`bulk' nuclear mean field.

On the other, a mechanism where a dynamical
deformation of the heavier partner of the reaction develops can also be recognized.
Eventually, this leads to the emission of the `attached neck' followed by a
Coulombian push (due to the proximity of the heavy source) that projects the
emitted particles towards the intermediate velocity range. This second
alternative occurs on a slower time scale and mainly involves the
emission of IMF particles. 

This second type of emission can be considered as a dynamically
induced asymmetric fission. In fact, the study of two reactions with an 
inverse asymmetry in the entrance channel allowed us to reinforce the idea 
that this kind of {\it delayed aligned emission} has its origin in a dynamical induced 
shape instability that mostly affects the heavier partner of the reaction. 
More important deformations can be sustained by the heaviest nucleus for enough
time to allow asymmetric fission processes to develop.


 Coulomb effects were analyzed in the $^{58}$Ni+Au
reaction, and no qualitative changes were observed in the
emission pattern of the QP, while major ones were reported for the QT.
This fact reinforces the idea that Coulomb instabilities develops mostly on
the side of the biggest partner, possibly not only because of the  
larger charge involved, but also because the dynamical
deformation settled by the reseparation dynamics is more pronounced on
the biggest nuclei side.

When the two partners of the reaction reseparate, the
biggest nucleus in an asymmetric reaction uses more of its surface 
nucleons than do the smallest one to establish the necklike bond between
the two poles. Therefore, the induced shape deformation on the biggest
nucleus after reseparation should be greater (the smallest nucleus has
relatively not enough surface to reabsorb the neck). The neck either
proceed through multiple neck ruptures before reabsortion, breaking up
in the reported aligned asymmetric way,  or is
effectively reabsorbed by the biggest nucleus.
Moreover, if we assume an equal energy sharing picture, valid for
peripheral events, the $Ni$ nucleus would be much hotter than the $Au$
one and could breakup by other processes before shape deformations
develops. Even if some fragments could still be emitted by dynamical deformations 
in this case, the associated partial cross sections would be very hard to
isolate.  
It is worth noting, though, that even if the presented picture favors the occurence of
IMF delayed emission on the biggest nucleus side, nothing forbids the phenomenon to appear 
on a less important scale on the smallest nucleus side.

\section{Conclusions}
In this paper, we have studied two asymmetric heavy ion reactions
using the description scheme of a classical molecular dynamics model.
We found that calculations are consistent with experiment, and that
processes described by the model could be relevant to the experimental case.

We have made
extensive use of the availability of the microscopic correlations at
all times provided by this approach.
Adopting the MSTE definition of {\it fragment}, the detailed knowledge
of the dynamics allowed us to  classify the complete set of asymptotic charged
particles according to its physical situation in phase space, right after
the collision (QP, QT or FP classification).
Correlating this information with the emission timescale associated to each
fragment, a
clear distinction between different modes of particle emission in the
mid-rapidity range was established. A fast emission
process that mainly involve LCP, and a delayed aligned emission of
IMF's, mainly coming from the heaviest partner of the reaction, were identified.

 The delayed aligned emission pattern reported in this
paper could possibly be compared to the statistical emission of reference
\cite{stat}, where a nucleus deexcite in the presence of the Coulomb field
of a secondary source. Results of Ref.\cite{stat} were obtained by
simulating a Au+Au reaction at 35 MeV/nucleon, therefore in the presence
of  strong Coulomb fields. According to our calculations, the same
phenomenon of delayed aligned emission can be traced even
in the $^{58}Ni$+C reaction, despite the small size of the TLF.
Experimental evidences for this effect were also reported in \cite{ging01}.
This could indicate that dynamical effects can not be disregarded in
the description of the process for light systems. 

All of these findings set a reasonable
framework in which recent experimental observations of particle
alignment and proximity effects of the heavier partner
can be understood \cite{ging01}. 
Moreover, they should be taken into account in the 
interpretation of other experimental observations. For example, in what concerns 
the geometrical and shape constraints that the observed delayed emission of
IMF's impose to chemical or thermal equilibration processes in nuclear
systems.

\section*{Acknowledgment}
This work has been supported by the University of
Buenos Aires (Grant No.TW98), CONICET (Grant No. PIP 4436/96),
the Natural Sciences and Engineering Research Council of Canada and
the Fonds pour la Formation de Chercheurs et l'Aide \`a la Recherche
du Qu\'ebec.
A.Ch. acknowledges CONICET for financial support and the warm
hospitality of the Laboratoire de Physique Nucl\'eaire, D\'epartement de 
Physique, Universit\'e Laval (Quebec, Canada).

\newpage

\begin{thebibliography}{99}
\bibitem{aclara}
We also took into account that the QP and QT must satisfy certain
kinematical constraints, for instance: after the collision the QP(QT) candidate
fragment should have a parallel velocity greater(lower) than zero in the 
center of mass reference system. For the mid-peripheral and peripheral reactions
analyzed in this paper this procedure seemed to be suitable enough to identify
reasonable quasi-sources.
\end{thebibliography}
\end{document}